\documentclass[conference]{IEEEtran}
\IEEEoverridecommandlockouts
\usepackage{cite}
\usepackage{amsmath,amssymb,amsfonts}
\usepackage{algorithmic}
\usepackage{graphicx}
\usepackage{textcomp}
\usepackage{tikz}

\usepackage{soul}
\def\BibTeX{{\rm B\kern-.05em{\sc i\kern-.025em b}\kern-.08em
    T\kern-.1667em\lower.7ex\hbox{E}\kern-.125emX}}

\usepackage{url}
\usepackage{makecell}
\usepackage{caption}
\usepackage{subfig}

\usepackage{float}

\usepackage{multirow}
\usepackage{lipsum}
\usepackage{caption}
\usepackage{wrapfig}

\usepackage{makecell}
\usepackage{xspace}
\usepackage{pifont}

\usepackage{booktabs}
\usepackage{tabularx}
\usepackage{ragged2e} 
\usepackage{array}      
\usepackage{lipsum}     

\newcolumntype{L}{>{\RaggedRight\arraybackslash}X}
\newcolumntype{C}[1]{>{\centering\arraybackslash}p{#1}} 

\usepackage{tikz}
\newcommand*\circled[1]{\tikz[baseline=(char.base)]{
            \node[shape=circle,fill,inner sep=1pt,scale=0.8] (char) {\textcolor{white}{#1}};}}

\usepackage{caption}
\usepackage{booktabs}

\usepackage{lipsum}
\usepackage{adjustbox}
\usepackage{bm}
\usepackage{soul}

{\color{red}}%
{}

\begin{document}

\title{
Characterizing and Optimizing LLM Inference Workloads on CPU-GPU Coupled Architectures\\
\thanks{\IEEEauthorrefmark{1}Authors were with Samsung Semiconductor when this work was done.}
\thanks{This work resulted from a CMU/Samsung collaboration  supported by Samsung's Global Research Outreach (GRO).}
}

\makeatletter
\newcommand{\linebreakand}{%
  \end{@IEEEauthorhalign}
  \hfill\mbox{}\par
  \mbox{}\hfill\begin{@IEEEauthorhalign}
}
\makeatother

\author{
  \IEEEauthorblockN{Prabhu Vellaisamy\IEEEauthorrefmark{1}}
  \IEEEauthorblockA{\textit{Carnegie Mellon University}\\
    pvellais@andrew.cmu.edu}
  \and
  \IEEEauthorblockN{Thomas Labonte}
  \IEEEauthorblockA{\textit{Samsung Semiconductor, Inc.}\\
    thomas.labonte.pub@gmail.com}
  \and
  \IEEEauthorblockN{Sourav Chakraborty}
  \IEEEauthorblockA{\textit{Samsung Semiconductor, Inc.}\\
    sourav.osu@gmail.com}
  \linebreakand 
  \IEEEauthorblockN{Matt Turner\IEEEauthorrefmark{1}}
  \IEEEauthorblockA{\textit{Hewlett Packard Enterprise}\\
    mattdturner@gmail.com}
  \and
  \IEEEauthorblockN{Samantika Sury\IEEEauthorrefmark{1}}
  \IEEEauthorblockA{\textit{Hewlett Packard Enterprise}\\
    samantika.sury@hpe.com}
  \and
  \IEEEauthorblockN{John Paul Shen}
  \IEEEauthorblockA{\textit{Carnegie Mellon University}\\
    jpshen@cmu.edu}
}

\maketitle

\begin{abstract}

Large language model (LLM)-based inference workloads increasingly dominate data center costs and resource utilization. Therefore, understanding the inference workload characteristics on evolving CPU-GPU coupled architectures is crucial for optimization. This paper presents an in-depth analysis of LLM inference behavior on loosely-coupled (PCIe A100/H100) and closely-coupled (GH200) systems. We analyze performance dynamics using fine-grained operator-to-kernel trace analysis, facilitated by our novel profiler \textit{SKIP} and metrics like Total Kernel Launch and Queuing Time (TKLQT). Results show that closely-coupled (CC) GH200 significantly outperforms loosely-coupled (LC) systems at large batch sizes, achieving 1.9x–2.7x faster prefill latency for Llama-3.2-1B. However, our analysis also reveals that GH200 remains CPU-bound up to 4x larger batch sizes than LC systems. In this extended CPU-bound region, we identify the performance characteristics of the Grace CPU as a key factor contributing to higher inference latency at low batch sizes on GH200. We demonstrate that TKLQT accurately identifies this CPU/GPU-bound transition point. Based on this analysis, we further show that kernel fusion offers significant potential to mitigate GH200’s low-batch latency bottleneck by reducing kernel launch overhead. This detailed kernel-level characterization provides critical insights for optimizing diverse CPU-GPU coupling strategies. This work is an initial effort, and we plan to explore other major AI/DL workloads that demand different degrees of CPU-GPU heterogeneous architectures.

\end{abstract}

\maketitle

\section{Introduction}

Heterogeneous CPU-GPU systems provide the essential infrastructure to support large-scale computing in modern data centers \cite{iyer2016heterogeneous, chen2016features, kant2009data}. Leveraging GPU advancements, these systems are enabling artificial intelligence (AI) breakthroughs. Generative AI (GenAI) applications, driven by large language models (LLMs), are transforming industries such as healthcare and e-commerce  \cite{uludag2023testing, sezgin2022operationalizing, kanbach2024genai, ghaffari2024generative}. As AI workloads proliferate, inference compute demand surpasses training's fixed cost \cite{desislavov2023trends, ai_inference}. Unlike training, inference scales with user demand, posing unique challenges for performance optimization and cost efficiency in data centers. Latency-sensitive inference workloads, such as autoregressive LLMs, highlight the need for performance-efficient system designs.

Recently, complex AI pipelines have emerged, such as the \textit{agentic} AI systems \cite{liang2023encouraging, du2023improving}, which consist of an LLM that orchestrates the coordinated behavior of multiple autonomous agents deployed for diverse applications. Another is \textit{Retrieval-Augmented Generation (RAG)} \cite{lewis2020retrieval}, a technique that combines information retrieval with text generation by performing a search over a knowledge base to obtain relevant context or facts to inform the subsequent generation process. These latency-sensitive applications necessitate optimizing processing unit (PU) interactions, especially for chained models or mixed CPU/GPU computations. Therefore, understanding the CPU-GPU interplay in emerging architectures is important for efficient LLM inference.

\begin{figure}[ht!]
    \centering
    \includegraphics[width=0.49\textwidth, height=4.3cm]{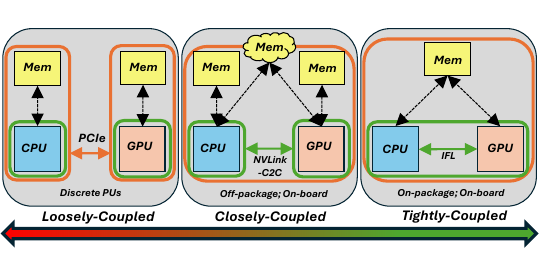}
    \caption{   
    \textit{\textbf{Evolution of CPU-GPU coupling paradigm}}. On the left, traditional data center architectures employ PCIe interconnects between discrete CPUs and GPUs, each maintaining separate memory pools. These systems are \textit{loosely-coupled} (LC). In the center, \textit{closely-coupled} (CC) architectures combine CPUs and GPUs on the same board, enabling unified memory access despite physically separate memories, and employ high-speed interconnects for data movement. On the right, \textit{tightly-coupled} (TC) architecture integrates the PUs within the same package and possesses a shared unified physical memory. 
    }
    \label{fig:coupling}
\end{figure}
    
Traditional data center architectures rely on PCIe interconnects connecting discrete CPUs and GPUs, referred to as \textit{loosely-coupled} (LC) architectures. These architectures suffer from high latency and limited interconnect bandwidth. Conversely, emerging \textit{tightly-coupled} (TC) architectures integrate heterogeneous PUs by sharing the same package (possibly with die stacking) and a shared unified physical memory and virtual address space like AMD's Instinct MI300A \cite{smith202411}. This trend towards tight-coupling marks a shift in how servers in data centers (Fig. \ref{fig:coupling}) are architected to meet the demands of modern AI workloads. In between these two architecture paradigms, the \textit{closely-coupled} (CC) architectures exist where heterogeneous PUs share the same board, utilize high-speed chip-to-chip interconnects, and employ unified virtual memory. Architectures like NVIDIA's Grace Hopper Superchip (GH200) fall under this broad category. We ask: \linebreak \circled{1} \textbf{\textit{Are CC/TC systems universally more effective for inference, or are LC systems more efficient in some scenarios?}}

Previous benchmarking studies on GH200 focus on data movement and performance/power efficiencies \cite{li2024automatic, simakov2024first, fusco2024understanding, hanawa2024preliminary} for various applications. However, the fine-grained dynamics of operator-to-kernel offloading and execution have yet to be investigated. Understanding these interactions, specifically how high-level deep learning (DL) framework operators translate into low-level GPU kernels, the associated CPU overheads, and GPU queuing/execution behavior, is crucial for optimizing system utilization as they dictate how efficiently kernels are offloaded, queued, and executed in heterogeneous systems. This paper seeks to characterize these fine-grained interactions. Hence, we address: \circled{2} \textbf{\textit{How do operator-kernel interactions evolve across these architectures, and what inefficiencies emerge?}}

Another aspect to consider is the impact of kernel fusion techniques on CC/TC architectures. Kernel fusion reduces kernel launch latencies and data movement overhead, increasing arithmetic compute intensity and enhancing inference performance \cite{yuan2024llm}. Techniques like FlashAttention \cite{dao2022flashattention, dao2023flashattention, shah2024flashattention} and CUDA Graphs \cite{cudagraph} consolidate operations into optimized kernels, minimizing launch counts and overhead. However, the benefits of fusion in CC/TC architectures, where interconnect bandwidth and memory access differ significantly, remain underexplored. \circled{3} \textit{\textbf{How do kernel fusion strategies benefit evolving CPU-GPU coupled systems for inference latency optimization?}}
    
To obtain these insights, we developed \textit{SKIP} (System-Aware Kernel Inference Profiler), a PyTorch-based profiler that builds operator-kernel dependency graphs from PyTorch Profiler \cite{pytorch_profiler} traces to benchmark LLM inference on state-of-the-art hardware. We propose novel metrics based on fine-grained operator-kernel offload data to evaluate system performance accurately. Using these metrics, particularly Total Kernel Launch and Queuing Time (TKLQT), we identify critical threshold regions for models transitioning from \textit{CPU-bound} region (characterized by GPU under-utilization) to \textit{GPU-bound} region (where kernel queuing and GPU stream utilization become the dominant factors). To address the inefficiencies of CPU-bound workloads across the evaluation platforms, we devise a new kernel fusion recommendation framework based on \textit{proximity score}. This metric quantifies deterministic kernel sequences, enabling data-driven recommendations to minimize the kernel launches, improve speedup, and balance PU utilization for CPU-bound workloads.

We list the main contributions of this work as follows:

\begin{enumerate}
     \item \textbf{Experimental benchmarking of LLM workloads on heterogeneous platforms}: Our experiments involve four LLM inference workloads and three CPU-GPU platforms. The four workloads include Bert-Base-Uncased, XLM-Roberta-Base, GPT2, and Llama-3.2-1B models. The three systems include two LC systems (PCIe-connected A100, H100 GPUs) and a CC GH200 system.

    \item \textbf{Tighter Coupling Trend}: While the trend towards tighter CPU-GPU coupling promises benefits, our work confirms this primarily for larger batch sizes where high memory bandwidth dominates. We also demonstrate that for latency-sensitive, low-batch workloads, the single-thread performance of the CPU remains a critical factor, potentially favoring powerful CPUs in LC systems.

    \item \textbf{TKLQT metric for characterizing PU-boundedness}: Among various new fine-grained operator-kernel metrics, we introduce the \textit{Total Kernel Launch and Queuing Time} (TKLQT) and demonstrate it as a more effective method to attribute compute-boundedness for LLM inference workloads, when compared to prior works.

    \item \textbf{CPU performance bottleneck for LLM inference}: We demonstrate that CPU performance can be a significant bottleneck for limiting low-batch latency. For example, we show that encoder-only models are 4x more CPU-bound for GH200 than for the two LC systems.

    \item \textbf{Effective regions for balanced utilization of PUs}: Each application-system pair exhibits optimal \textit{sweet spot} batch sizes. In this balanced region, both CPU and GPU are effectively utilized. Identifying and operating within this region is key for maximizing overall system efficiency rather than targeting maximum GPU saturation.

    \item \textbf{Proximity-based kernel fusion recommendation}: We develop a novel and general kernel fusion recommendation method based on kernel proximity score, enabling CPU-bound inference workloads, particularly prevalent in CC systems at smaller batch sizes, to transition towards more balanced CPU-GPU utilization.

    \item \textbf{SKIP profiling tool}: We developed a novel PyTorch-based operator-kernel profiling tool, SKIP, to enable evaluations of diverse heterogeneous platforms. SKIP is also used to perform kernel fusion recommendation analysis.

\end{enumerate}

\section{Background}
\label{sec:background}

This section discusses the background topics related to this work, including emerging latency-sensitive inference workloads, different CPU-GPU coupling paradigms, kernel fusion techniques, current profiling tools, and comparisons with prior related works. 

\subsection{LLM Workloads and Emerging Datacenter Applications}

AI inference costs increasingly exceed training costs, with potentially up to 90\% AI-related costs attributed to inference due to its repeated execution across millions of user requests in pervasive AI systems \cite{desislavov2023trends}. The Transformer architecture \cite{vaswani2017attention} has become the dominant framework for natural language generation (NLG) tasks, including text summarization, machine translation, and question answering, with encoder-only and decoder-only models commonly employed. This work focuses on LLM inference, specifically encoder-only and decoder-only workloads, to derive insights for optimizing latency in evolving CPU-GPU coupled systems. 

A key challenge with LLM inference is the variability of workloads due to the diverse nature of requests. The request range can be short (a few tokens) to long (a hundred thousand tokens), with similar variability for output generation. Longer inputs necessitate increased GPU parallelism, resulting in extended \textit{prefill} phases, while longer outputs induce multiple iterations and elongated \textit{decode} phases \cite{stojkovic2024towards}. The prefill stage puts pressure on the compute resources, while the decode stage puts pressure on the memory subsystems \cite{patel2023polca}\cite{patel2024splitwise}.

Batch size selection profoundly impacts the user experience and perceived responsiveness in real-time AI applications. Increasing the batch size could linearly worsen the latency of LLM inference \cite{agrawal2024taming, yu2022orca}. System-level objectives constrain the latency to around 200 ms to ensure a good user experience \cite{choi2021lazy, shen2019nexus}. While large batch sizes provide high throughput and efficiently handle large volumes of data, low-batched inference ensures immediate feedback and low latency. 

In agentic AI systems, an LLM orchestrates the coordinated behavior of multiple autonomous agents deployed for diverse applications. These systems generate output that serves as input to downstream models or processes, enabling complex task execution and decision-making where a single model orchestrates multiple models and data sources. With conversational AI agents \cite{wu2023mathchat} and multi-agent AI systems \cite{wu2023autogen}, the complexity of the pipeline increases but still requires a fast response to the user. If each component within the system increases latency due to larger batch sizes, the cumulative delay can become substantial, leading to an unacceptable overall response time. 

In RAGs \cite{lewis2020retrieval}, during the final generation phase (consisting of LLM), processing in large batch sizes can improve LLM throughput but can increase latency for individual users, especially for a time-to-first token (TTFT). As AI models evolve and the complexity of these pipelines increases, the chaining of outputs and inputs across multiple models is expected to become more prevalent. This increase in pipeline complexity will further drive the need for optimization techniques focused on minimizing latency, particularly between the CPU and GPU, when computation is split between both.

\subsection{CPU-GPU Coupling Paradigms}

Heterogeneous CPU-GPU architectures are foundational for diverse data center workloads. Usually, CPUs are responsible for control flow and orchestration, while GPUs accelerate parallel tasks such as matrix computations and DL workloads. Traditional data center LC architectures use PCIe interconnects, with separate CPU/GPU memory pools. This architecture scales well and accommodates PU flexibility, but LC suffers latency and bandwidth limits with larger model sizes.

Data centers are migrating towards integrating tighter coupled architectures to address data transfer-related bottlenecks. For example, NVIDIA's GH200 employs NVLink-C2C (Chip-to-Chip) \cite{wei20239}, a high-speed, low-latency interconnect, offering up to 900 GB/s of bidirectional bandwidth (approximately 7x faster than PCIe Gen5 lanes \cite{fusco2024understanding}) which unifies memory accesses between PUs, sharing a unified virtual address space. Furthermore, the CPU and GPU packages are integrated onto the same board to minimize NVLink-C2C interface length. The NVLink-C2C interconnect exposes the LPDDR5X memory of the CPU and the HBM3 memory of the GPU as separate domains of non-uniform memory access (NUMA), helping mitigate data transfer limitations by allowing direct access to memory without full duplication.

Following this trend towards even tighter CPU-GPU coupling is the AMD MI300A architecture, which combines AMD Zen4 x86 CPU cores with the CDNA3 GPU units, with unified HBM3 shared coherently between them, all integrated into the same package by stacking chips on a single substrate. Each APU provides 1 TB/s of bidirectional connectivity through eight 128 GB/s AMD Infinity Fabric interfaces \cite{smith202411}. This contrasts with the GH200 architecture, where memory is unified only virtually. MI300A benefits from the absence of explicit CPU-GPU data transfer due to its physically unified memory architecture. This allows for tighter data coherence and more efficient memory access. The GH200 and MI300A architectures represent significant steps forward in mitigating the data transfer bottlenecks inherent in heterogeneous systems, but their operator-kernel interactions, contrasting to LC systems, are experimentally unexplored.

\begin{figure}
    \centering
    \includegraphics[width=0.45\textwidth]{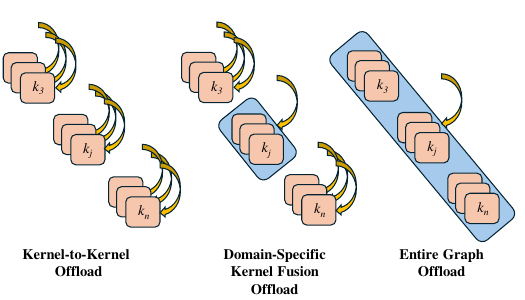}
    \caption{\textbf{\textit{Types of Kernel Fusion}}. The figure depicts kernel sequence $k_1, k_2, k_3,...,k_n$ in a GPU stream, triggered by CPU operators. From left to right: (1) Kernel-to-kernel offload (eager mode, unfused), (2) Domain-specific operator fusion (e.g., FlashAttention fusing self-attention operators), and (3) Entire graph offload (e.g. torch.compile/CUDA Graphs fusing larger subgraphs/whole graph).}
    \label{fig:types_fusion}
\end{figure}

\subsection{Kernel Fusion Techniques}

To accelerate computational tasks in GPUs, especially for AI, most DL frameworks support various kernel fusion techniques \cite{wang2010kernel, fousek2011automatic, wahib2014scalable, lin2022building}. A kernel is a tiny parallelizable subtask in a GPU computational task dispatched to the GPU for execution. In kernel fusion, two or more of the subtasks are combined into a larger and equivalent kernel to improve performance by mitigating data round trips to GPU memory, making better use of GPU resources \cite{lin2022building} and achieving energy improvements \cite{wang2010kernel}. This work analyzes kernel fusion benefits solely through reduced kernel launch counts. 

Frameworks like PyTorch \cite{ketkar2021introduction} offer these techniques as different execution modes for inference (Fig. \ref{fig:types_fusion}). Traditionally, frameworks perform eager-mode execution, where kernels are launched and executed as encountered in the operators. This mode does not incur compilation overhead and is more flexible than kernel fusion techniques. However, they can suffer from an increased kernel launch tax (fixed cost of launching a kernel when GPU is not fully utilized) due to multiple kernel launches. On the other hand, kernel fusion techniques fuse kernels to be launched together, saving the kernel tax overhead incurred from multiple individual kernel launches. Some PyTorch-specific kernel fusion techniques analyzed in this work are as follows.

\begin{itemize}

\item \textit{Domain-specific}: Techniques such as FlashAttention fuse multiple domain-specific operations into a single kernel, reducing memory access, and improving efficiency in attention-based models. It limits multiple accesses to the HBM in GPU and instead performs an iterative-based fine-grained attention mechanism to fill the local cache \cite{dao2022flashattention}, mitigating IO-related transfer overhead. In terms of FlashAttention operator-kernel relationships, it fuses specific kernels related to self-attention operation into one kernel to perform its execution and is an example of domain-specific kernel offload. In this work, we use FlashAttention2 \cite{dao2023flashattention} for evaluation.

\item \textit{Graph Synthesis (torch.compile)}: With torch.compile \cite{torch_compile}, PyTorch further optimizes execution by tracing and compiling computational graphs in the form of CUDA graphs \cite{cudagraph}. The entire network can be synthesized into a single computation graph, launched only once at the beginning, resulting in reduced synchronization and launch latency overheads but at the cost of increased compilation time for graph synthesizing. 
Table \ref{tab:torch_compile_modes} reports the compilation time overhead and resultant speedups for various torch.compile modes relative to the conventional eager mode execution for the Gemma-2B model on the Intel CPU+H100 evaluation platform (specification in Table \ref{tab:system-specs}). As reported, the compilation time overhead can rise from 15.3x to 944.6x increase when compared to eager mode execution, with limited flexibility, such as not accommodating resizing KV (key-value) cache dynamically and requiring recompilation for changes in batch sizes.
Fig. \ref{fig:flash_compile_speedups} reports the speedup comparisons achieved using domain-specific kernel fusions like FlashAttention2 and torch.compile max-autotune mode over eager mode execution for popular 7B LLMs on the same platform. Max-autotune involves backend compilers like Triton \cite{tillet2019triton} to boost performance through optimized kernels and graph synthesis. In this evaluation, we only considered fully synthesized inference graphs. 
\end{itemize}

\begin{table}[!ht]
    \centering
    \renewcommand{\arraystretch}{1.2} 
    \setlength{\tabcolsep}{4pt} 
    \caption{TTFT (time-to-first-token) compilations and speedups for various torch.compile modes relative to eager mode execution for the Gemma-2B model with batch size of 1 and input sequence of 1024 tokens. Evaluation platform is Intel Xeon Platinum connected to NVIDIA H100 over PCIe Gen5.}
    \label{tab:torch_compile_modes}
    \begin{tabular}{|c|c|c|c|c|}
        \hline
        \textbf{Compile Mode} & 
        \makecell[c]{\textbf{Eager}} & 
        \makecell[c]{\textbf{Default}} & 
        \makecell[c]{\textbf{Reduce-} \\ \textbf{overhead}} & 
        \makecell[c]{\textbf{Max-} \\ \textbf{autotune}} \\ \hline
        \textbf{Compilation Time (s)} & 0.40644 & 6.2844 & 12.7469 & 387.3 \\ \hline
        \textbf{Speedup} & 1 & 1.203 & 1.2394 & 1.317 \\ \hline
    \end{tabular}
\end{table}

\begin{figure}[h!]
    \centering
    \includegraphics[width=0.45\textwidth, height=3.5cm]{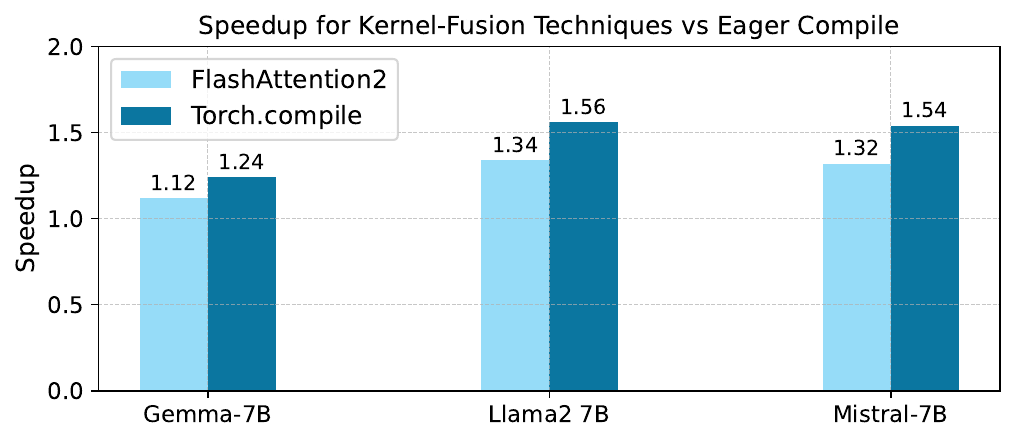}
    \caption{TTFT speedups for FlashAttention-2 and torch.compile max-autotune mode for various 7B decoder models (compared to eager mode execution). Evaluation platform is Intel Xeon Platinum connected to NVIDIA H100 over PCIe Gen5. }
    \label{fig:flash_compile_speedups}
\end{figure}


\subsection{Profiling Tools and Prior Works}


Industry-grade profiling tools provide insight into AI workload performance in heterogeneous architectures, but often have limitations in analyzing complex kernel interactions and offload characteristics in large-scale models. NVIDIA Nsight Systems \cite{nsight_systems} identifies bottlenecks, latencies, and resource utilization, but lacks visibility into the PyTorch Aten operators on the CPU, which requires expert intervention. Similarly, Nsight Compute \cite{nsight_compute} performs intra-kernel analysis of metrics like memory bandwidth and warp efficiency, but is limited to GPU-specific profiling and fails to capture offload interactions between CPU and GPU. PyTorch Profiler \cite{pytorch_profiler} offers profiling of CPU and CUDA kernel executions, capturing useful metrics for model optimization through CUPTI \cite{nvidia_cupti} to collect low-level CUDA performance metrics. SKIP extends this by constructing detailed operator-kernel dependency graphs and calculating specific metrics like TKLQT, enabling a deeper analysis of launch overheads and queuing dynamics.


Previous academic work focuses mainly on optimizing DNN training. Daydream \cite{zhu2020daydream} simulates model execution using dependency graphs built from low-level CUPTI traces, introducing graph transformation primitives to model various optimizations. ASTRA-Sim \cite{rashidi2020astra} enables parameterized simulations of DNN training for an end-to-end study of communication bottlenecks in distributed DNN training. Other optimization strategies operate at the computation graph level before execution. TASO \cite{jia2019taso} automatically generates graph substitutions to rewrite DNN operator graphs into more efficient forms, enabling better operator fusion by backend compilers like XLA or TensorRT. dPRO \cite{hu2022dpro} collects communication traces across TensorRT and MXNet to predict DNN distributed training performance, and similarly optimizes operator graphs through operator fusion. Thes prior works contrast with this work, which analyzes fine-grained runtime traces at the kernel level to analyze kernel-level metrics and recommend kernel fusions based on kernel execution proximity.

Recent work \cite{fernandez2023framework} explored execution bounds based on whether models are constrained by ML framework tax or GPU kernel compute (\textit{framework-bound vs. compute-bound}) by observing end-to-end latency scaling with respect to batch size. We arrive at a similar classification, \textit{CPU-bound vs. GPU-bound} by a different, more fine-grained metric derived from kernel launch and queue times. With \cite{fernandez2023framework}, the flat latency curve implies framework overhead dominance but does not directly measure which overhead or its magnitude. Compared to \cite{fernandez2023framework}, our classification through total kernel launch and queuing time (TKLQT) pinpoints inefficiencies directly related to the CPU-GPU interaction and GPU saturation (see Section \ref{sec:boundedness}).

These gaps motivate our development of SKIP, which specifically analyzes operator-kernel offload dynamics during inference, as detailed in this work. The closest comparable tool, TraceSim \cite{liang2024fine}, utilizes PyTorch Kineto to extract meta-information for simulating `what-if' scenarios and performance predictions for training workloads. Prior kernel profiling studies are typically limited to specific applications and older GPU architectures \cite{lin2022building, lin2023performance, liang2024fine}. Table \ref{tab:tools} provides a comparative summary of related works against this work.


\begin{table*}[t] 
\caption{Comparisons with Previous Works}
\label{tab:tools}
\centering
\footnotesize
\begin{tabularx}{\textwidth}{@{} l L C{1.3cm} C{1.1cm} C{1.3cm} C{1.6cm} C{2 cm} @{}} 
\toprule
\textbf{Tool/Method} & \textbf{Primary Focus} & \textbf{CPU-GPU Analysis} & \textbf{Fusion Support} & \textbf{Exec. Mode} & \textbf{Target Workloads} & \textbf{Evaluation Platform(s)} \\ 
\midrule
Daydream [43] & Predicting training optimization impact via simulation on kernel dependency graphs. & No & No & Training & ResNet, BERT & NVIDIA 2080Ti \\

ASTRA-Sim [29] & Network simulation for distributed training (topology, collectives, SW/HW co-design). & No & No & Training & Large-scale DNNs & Simulation  \\

dPRO [16] & Diagnosis \& optimization (including operator fusion) of distributed training using operator \& fine-grain comm. traces. & No & Yes (Op/ Tensor) & Training & MXNet / TensorRT & NVIDIA V100 \\

Fernandez et al. [10] & Observing/defining framework tax based on inference latency scaling. & No & No & Inference & NLP models & Various NVIDIA GPUs (Pascal, Voltus, and Ampere) \\

TraceSim [24] & Predictive simulation of training performance ('what-if', scaling) using kernel traces. & Partial (simplified CPU profiling) & No & Training & Vision / Transformers & NVIDIA H100 Cluster \\ 
\midrule
\textbf{This Work (SKIP)} & \textbf{Diagnoses inference bottlenecks (CPU/GPU bound, launch tax) \& recommends kernel fusion from kernel traces.} & \textbf{Yes} & \textbf{Yes (kernels)} & \textbf{Inference} & \textbf{LLMs (BERT, GPT2, Llama)} & \textbf{NVIDIA A100, H100, GH200} \\
\bottomrule
\end{tabularx}
\end{table*}

\begin{figure}[ht!]
    \centering
    \includegraphics[width=0.5\textwidth, height= 3.1cm]{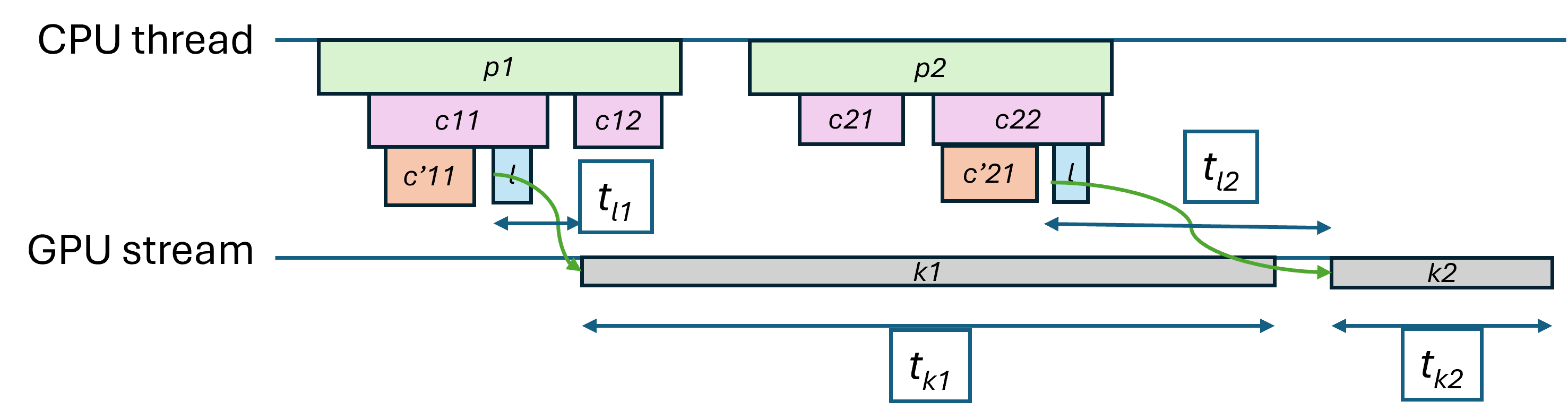}
    \caption{Operator-kernel execution timing. Illustrates how CPU-side operators trigger GPU kernel execution, with the launch latency ($t_l$) illustrated as the latency between start of CPU launch call $l$ to start of execution of kernel $k$.}
    \label{fig:kernel_metrics}
\end{figure}

\section{Proposed System Performance Metrics and Kernel Fusion Recommendation}
\label{sec:skip}

This section proposes performance metrics derived from SKIP's fine-grained analysis of LLM inference traces. We further describe how using the Total Kernel Launch and Queuing Time (TKLQT) metric contributes towards a novel classification of PU-boundedness. Consequently, a high-level overview of the new kernel fusion method using proximity score.

\subsection{Fine-Grained Kernel Analysis}

Unlike previous tools, SKIP provides a finer-grain view of operator-kernel offloads, crucial for analyzing offload patterns in AI inference. Based on the stream trace (example illustration depicted in Fig. \ref{fig:kernel_metrics}), SKIP derives operator-kernel relationships by creating a dependency graph between the operators and kernel timestamps. The key metrics derived are calculated as follows.

\subsubsection{Total Kernel Launch and Queuing Time (TKLQT)} 
Assume an operator $o_i$ consists of a parent ATen operator ($p_i$) for the child operators ($c_{i1}, c_{i2}$, ... $c_{iN}$). Let $c_{i1}$ have a subsequent child operator ($c'_{i1}$) and a CPU launch call (cudaLaunchKernel) $l_i$, which triggers execution of a kernel $k_i$. Then, kernel launch overhead ($t_{li}$) is defined as:
\begin{equation}
    t_{li} = ts_b(k_i) - ts_b(l_i)
\end{equation}
where $ts_b$ is the beginning time stamp in the profiled traces. 
This duration includes the CPU time for the launch call, potential driver overhead, and transfer time (if data is implicitly moved). If there are intervening GPU kernels finishing compute while the operator launches the kernel, $t_l$ extends to kernel launch and queuing time. The total kernel launch and queuing time (TKLQT) is defined as: 
\begin{equation}
    TKLQT = \sum_{j=1}^{n} t_{lj}
\end{equation}
where $n$ is the total number of kernels launched. TKLQT directly measures the accumulated overhead associated with the CPU-GPU kernel offload process, making it highly sensitive to CPU limitations (launch cost) and GPU saturation (queuing). This metric is utilized attributing PU-boundedness for different workloads.

\subsubsection{Average Kernel Duration (AKD)}
For a kernel ($k_i$), the duration of the processing is defined as $t_{ki}$. The average kernel duration (AKD) is specified as:
\begin{equation}
    AKD = \frac{\sum_{j=1}^{n} t_{k_j}}{n}
\end{equation}
This metric indicates the inference computational intensity. Smaller AKD means lightweight or highly optimized kernels, while longer AKD suggests more computationally intensive kernels. It also conveys \textit{kernel efficiency}.

\begin{figure}[h!]
    \centering
    \includegraphics[width=\columnwidth, height = 4cm]{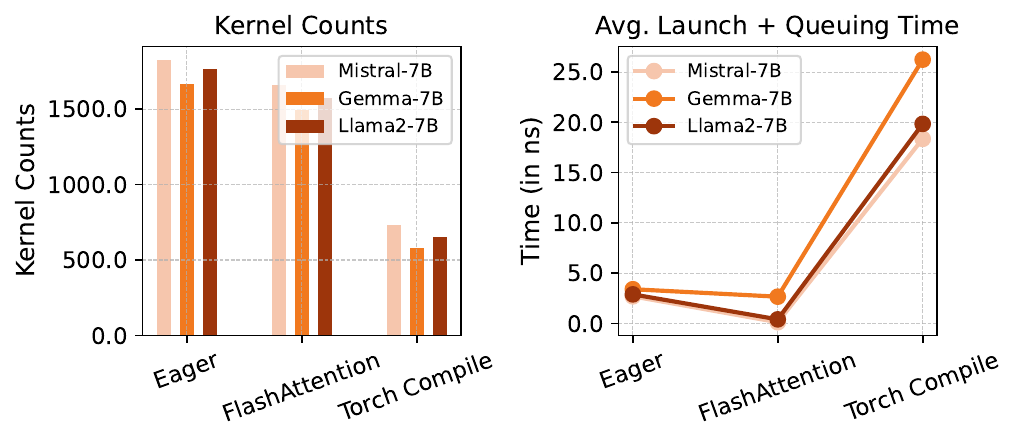}
    \caption{Increasing computation in each GPU kernel allows the CPU to queue up many kernels while hiding the offload latency. However, too much GPU computation results in large CPU idle times and long latencies to batch completion, translating to user-visible response latency.}
    \label{fig:}
\end{figure}

\subsubsection{Inference Latency (IL)}
For inference workload launching $n$ kernels in  execution, inference latency is defined as 
\begin{equation}
    IL = ts_e(k_n) - ts_b(p_1) 
\end{equation}
where $ts_e(k_n)$ is the ending timestamp of the last executed kernel and $p_1$ is the first parent Aten operator.

\subsubsection{GPU Idle Time}
We calculate the GPU idle time by subtracting the total kernel execution duration from the inference latency (IL).
\begin{equation}
    \textit{\text{GPU Idle Time}} = IL -  \sum_{j=1}^{n} t_{k_j}
\end{equation}
\subsubsection{Top-\textit{k} Kernel Tracking}
Through top-\textit{k} kernel tracking, SKIP pinpoints the most frequently invoked kernels, facilitating focused analysis on those with the highest offload tax. This allows micro benchmarking of specific kernels, aiding in quantifying kernel offload tax across architectures.

\subsection{Workload Classification Based on PU-Boundedness}

The goal is to optimize performance by minimizing user-visible latency (maximizing tokens processed per second for individual users) while maximizing CPU and GPU utilization by reducing idle time. Achieving this balance requires careful adjustment of the batch sizes. Larger batch sizes are often used to fully utilize GPU resources and improve throughput, but they can introduce delays in token generation for individual users, thereby increasing user-visible latency. Excessive large batch sizes may also lead to significant CPU idle time as the CPU waits for GPU computations to complete. Striking the right balance between batch size and resource utilization is crucial to maintaining high throughput without compromising the system's responsiveness to end users.

By tracking kernel launch and execution patterns, we can enable the classification of AI models as CPU-bound or GPU-bound. Accurate categorization of model characteristics is devised by identifying the conditions under which models transition from CPU-bound (low GPU utilization) to GPU-bound (intensive kernel queuing) and provides a foundation for targeted optimizations. We utilize the TKLQT metric to classify and provide insights into workload behavior as batch size increases, highlighting the inflection point at which workload transitions from CPU-dominated to GPU-intensive. 

Previous work \cite{fernandez2023framework} classified models based on either framework-bound or compute-bound, primarily focusing on inference latency and the impact of increasing batch sizes. However, the framework tax includes the sum of all non-GPU compute time, including Python interpretation, graph building, and more. TKLQT isolates a critical part of that tax, specifically, the latency inherent in the kernel offload process. TKLQT pinpoints inefficiencies directly related to the CPU-GPU interaction and GPU saturation by focusing on the launch-to-execution interval. This is often the most relevant part of the ``tax" to optimize when improving GPU utilization or reducing launch overhead. Further, TKLQT allows attributing performance differences in the CPU-bound region more directly to the CPU/launch subsystem, while from \cite{fernandez2023framework}, everything is encapsulated under the framework tax umbrella. Hence, our method is more favorable when evaluating compute efficiencies in coupled architectures and provides unique insight into architectural recommendations for future CC and TC architectures.

\subsection{Kernel Fusion Recommendations Using Proximity Score} 
We further integrate SKIP with scalable kernel-fusion recommendations based on derived inference traces. It outputs a list of kernel sequences recommended for fusion based on runtime traces from SKIP containing timed kernel sequences. We define proximity score (PS) as the likelihood of a chain of kernels of length $L$ occurring every time a kernel $k$ is executed. These sequences represent fixed computational patterns executed repeatedly, and fusing them amortizes the launch cost associated with each kernel in the sequence over a single launch, in turn directly reducing the TKLQT bottleneck identified in the CPU-bound region. Specifically, for kernel execution sequences $S_1 = [k_{1_1}, k_{1_2}, k_{1_3}, ..., k_{1_{n_1}}]$, $S_2 = [k_{2_1}, k_{2_2}, k_{2_3}, ..., k_{2_{n_2}}]$, ..., $S_N = [k_{N_1}, k_{N_2}, k_{N_3}, ..., k_{N_{n_N}}]$ separated by intervening CPU operator dependency, let a kernel chain $C$ of length $L$ be defined as $C = (k_i, k_{i+1}, k_{i+2}, ..., k_{i+L-1})$. Then, proximity score for $C$ is defined as 
\begin{equation}
    PS(C) = \frac{f(C)}{f(k_i)}
\end{equation}
where $f(C)$ is the frequency of chain $C$, while $f(k_i)$ is the frequency of the first kernel ($k_i$) in the chain $C$. To recommend fusion based on a proximity score threshold $T$, we suggest $PS(C) \geq T$, where $T$ is the minimum required PS for chain $C$ to be considered a candidate for fusion. If $PS(C)=1$, it signifies a deterministic pattern that is ideal for kernel fusion. 

Unlike domain-specific fusions like FlashAttention or full graph synthesis approaches like torch.compile, which require pre-specification or graph capture, our proximity score method automatically identifies general, deterministic kernel sequences directly from runtime traces, offering a flexible way to target launch overhead for frequently occurring patterns. Identifying fusion opportunities in the CPU-bound region where reducing launch tax (and thus TKLQT) has the most impact on overall latency, with relatively little to no speedup for GPU-bound region due to kernel queuing being dominant. In this work, only recommendations for fusion are proposed. Implementation using kernel compilers (e.g., Triton) or manual coding is planned for future work.

\section{SKIP Profiling Tool and Experimental Methodology}

In this section, we describe the SKIP profiling tool and the evaluation methodology used for this work.

\subsection{System-Aware Kernel Inference Profiler (SKIP)}

SKIP is built on PyTorch, enabling it to leverage the framework to run LLM inference and generate PyTorch Profiler, which uses CUPTI to capture GPU kernel events with high fidelity. It builds dependency graphs spanning parent PyTorch Aten operators, multiple child operators, CUDA runtime calls, and finally, kernels. SKIP processes the timestamped events from PyTorch Profiler traces. An Aten operator $p$ is designated as the parent of a subsequent child operator $c$ and/or CUDA runtime call $l$, if their start times fall within $p$'s duration ($tsb(p)$ to $tse(p)$), where $tse(p)$ is the end timestamp of $p$. Kernels $\mathbf{k}$ are linked to their corresponding launch call $l$ based on the CUDA correlation IDs provided by CUPTI. Using this process, SKIP is able to create accurate dependency mapping for all operator-kernel relationships by incorporating the stream and thread information embedded in the traces. Hence, SKIP accurately models the execution flow and calculates metrics such as TKLQT and AKD.

\subsection{Evaluation Method: AI Workloads and Platforms Used}

For our experimental evaluation of AI inference workloads, we benchmark the LLM models listed in Table \ref{tab:inf-workloads}. The pre-trained models are imported from the Huggingface library. All models used for evaluation are FP16 precision-based PyTorch models to ensure FlashAttention compatibility and fair comparisons. Table \ref{tab:system-specs} lists the evaluation platforms used. We evaluate CUDA platforms due to mature support through PyTorch. We use Python 3.10.12 and PyTorch 2.4.1 versions for all platforms, and CUDA 12.6 and NVIDIA driver 560.35.03 to keep software compatibility for comparisons. 

Experiments are conducted across various batch sizes to demonstrate kernel-level metric scaling. A batch size of 1 (BS=1) represents a latency-critical scenario (benchmarked by MLPerf Inference SingleStream \cite{mlperf}) such as chatbots \cite{jacoby2024human}, conversational AI agents \cite{wu2023autogen, yao2023react}. Further, serving frameworks like vLLM \cite{kwon2023efficient} aim to maximize throughput while approaching the low latency characteristic of BS=1 execution, highlighting its importance for interactive services, using PagedAttention and continuous batching. Smaller batch sizes (e.g., 4, 8, 16) reflect potential near real-time scenarios or high-concurrency serving architectures. Larger batch sizes (32+) represent throughput-oriented tasks and allow us to observe the transition into the GPU-bound region (depending on model complexity), providing insights into maximum hardware utilization. Unless otherwise specified, the benchmarks used a consistent input sequence length of 512 tokens to generate prefill latency (TTFT). Baseline results are obtained using PyTorch eager mode execution, unless compared with fused modes like torch.compile or FlashAttention2.


\begin{table}[!ht]
    \centering
    \renewcommand{\arraystretch}{1.3} 
    \setlength{\tabcolsep}{8pt} 
    \caption{LLM models used for workload benchmarking in our experimental evaluations.}
    \label{tab:inf-workloads}
    \begin{tabular}{|c|c|}
        \hline
        \textbf{Type of LLM} & \textbf{Inference Workload (parameters)} \\ \hline
        Encoder-only & 1. Bert-Base-Uncased (110M) \\ 
                     & 2. XLM-Roberta-Base (279M) \\ \hline
        Decoder-only & 1. GPT2 (137M) \\
                     & 2. Llama-3.2-1B (1.24B) \\ \hline
    \end{tabular}
\end{table}

\begin{table}[!ht]
    \centering
    \renewcommand{\arraystretch}{1.3} 
    \setlength{\tabcolsep}{4pt} 
    \caption{System specifications of CPU-GPU coupled platforms used for experimental evaluations in this study.}
    \label{tab:system-specs}
    \begin{tabular}{|c|p{6.2cm}|}
        \hline
        \multicolumn{1}{|c|}{\textbf{Type of Coupling}} & \multicolumn{1}{c|}{\textbf{System Specification}} \\ \hline
        Loosely-Coupled & 1. AMD EPYC 7313 16-Core Processor with 512GB, A100-SXM4-80GB (500W): AMD+A100\\
                        & 2. 2P Intel Xeon Platinum 8468V (48-core) with 512GB, H100 PCIe (350W): Intel+H100\\ \hline
        Closely-Coupled & Nvidia Grace Hopper Superchip (GH200): 480GB LPDDR5X Grace 72-core Arm Neoverse V2, 96GB HBM3 (900W) H100 \\ \hline
    \end{tabular}
\end{table}

\section{Experimental Evaluation \& Results}
\label{sec:results}
\subsection{Benchmark of nullKernel Launch Latency}

We benchmark nullKernel launch overhead on the three evaluation platforms to expose the underlying platform-level overheads from memory management and thread scheduling. We report the results in Table \ref{tab:nullkernel}. While not directly critical for inference latency optimization, these metrics provide essential insights into architectural trade-offs and help contextualize workload efficiency. This measured nullKernel overhead represents a baseline component of the TKLQT, which was analyzed later and reflected fixed costs independent of GPU load. It is particularly relevant when the system is CPU-bound and not experiencing queuing delays. The GH200 platform incurs slightly higher kernel launch overhead (roughly 400-500 ns more than other platforms). This could be due to lower single-core CPU performance on the GH200 or possibly due to its unified virtual memory management. It also has the lowest nullKernel execution durations. In contrast, the AMD+A100 platform exhibits the least launch overhead but faces the highest kernel execution durations. The Intel+H100 strikes a balance in offering moderate launch overhead and execution times. Although the GH200 introduces minor delays, it improves execution efficiency, making it suitable for workloads that benefit from tightly integrated memory systems.

\begin{table}[!ht]
    \centering
    \renewcommand{\arraystretch}{1.2} 
    \setlength{\tabcolsep}{4pt} 
    \caption{CudaLaunchAPI with nullKernel launch overhead and nullKernel duration across the evaluation platforms.}
    \label{tab:nullkernel}
    \begin{tabular}{|c|c|c|}
        \hline
        \textbf{Platforms} & \textbf{nullKernel Launch} & \textbf{nullKernel Duration} \\
                          & \textbf{Overhead (ns)}      & \textbf{(ns)}                \\ \hline
        \textbf{AMD+A100}   & 2260.5 & 1440.0 \\ \hline
        \textbf{Intel+H100} & 2374.6 & 1235.2 \\ \hline
        \textbf{GH200}      & 2771.6 & 1171.2 \\ \hline
    \end{tabular}
\end{table}

\textbf{Key Takeaway: \textit{The nullKernel launch latency evaluation reveals sensitivity to platform-level differences in CPU performance, GPU performance, memory management, and scheduling. This baseline launch overhead contributes directly to inference latency in CPU-bound scenarios. GH200 shows slightly higher overhead (2771.6 ns) than AMD+A100 (2260.5 ns) and Intel + H100 (2374.6 ns), but compensates with the fastest execution times, making it ideal for GPU-dependent workloads. Both LC systems here have a more powerful CPU, which can minimize launch overhead, favoring latency-sensitive tasks.}}

\subsection{Classifying CPU-bound vs GPU-bound for Workloads }
\label{sec:boundedness}

To optimize AI/LLM inference workloads, it is critical to understand whether a model’s performance is limited by CPU or GPU resources. Using SKIP, we introduce a novel method to classify workloads as CPU-bound or GPU-bound based on TKLQT. Experiments (Fig. \ref{fig:encoder_classification}) show constant TKLQT at small batch sizes, indicating little to no kernel queuing and GPU under-utilization. Hence, TKLQT in this region is due to pure kernel launch overheads. We term this region as CPU-bound, where the latency is bounded by CPU execution. The inflection points mark (star markers) the transition from a CPU-bound region, where TKLQT is dominated by kernel launch overheads, to a GPU-bound region, where kernel queuing becomes the primary contributor to TKLQT. 
As seen in Fig. \ref{fig:encoder_classification}, for encoder-only models, this transition occurs around a batch size of 8 for the LC systems (Intel+H100, AMD+A100), but is delayed significantly to around batch size 32 for the CC GH200, resulting in a 4x more CPU-bound region. Since the compute capabilities of the H100 and the GPU portion of the GH200 are equivalent, the difference between the two is likely due to the higher bandwidth HBM of the GH200. Because of this, the GPU of the GH200 can complete more work in the same amount of time as the H100, allowing a larger batch size to be completed within the shadow of CPU execution. Therefore, in this CPU-bound region, TKLQT reflects the aggregation of per-kernel launch costs (nullKernel launch taxes) plus minimal queuing delay.

Beyond the inflection point, kernel queuing in the GPU becomes the dominant factor over kernel launch overhead, contributing to high TKLQT. The CPU continuously queues up kernels on the GPU while the GPU remains fully utilized. Minimizing kernel launches through kernel fusion in this GPU-bound region provides no benefit. Hence, as discussed in the next subsection, SKIP's kernel fusion recommendation through the novel method of proximity score is targeted for CPU-bound workloads. Such kernel launch minimization techniques are more amenable on the CC GH200 system, considering that they delay the PU-boundedness by 4x compared to the other platforms in the case of encoder-only models.

\textbf{Key Takeaway: \textit{Workload classification using TKLQT reveals the critical transition points from CPU-bound to GPU-bound regions. GH200 delays this transition to larger batch sizes, leveraging high-bandwidth memory, 4x larger compared to the Intel+H100 and AMD+A100 systems. Hence, latency optimizations through kernel launch minimization are very amenable for CC/TC systems.}}

\begin{figure}[]
    \centering
    \includegraphics[width=0.5\textwidth]{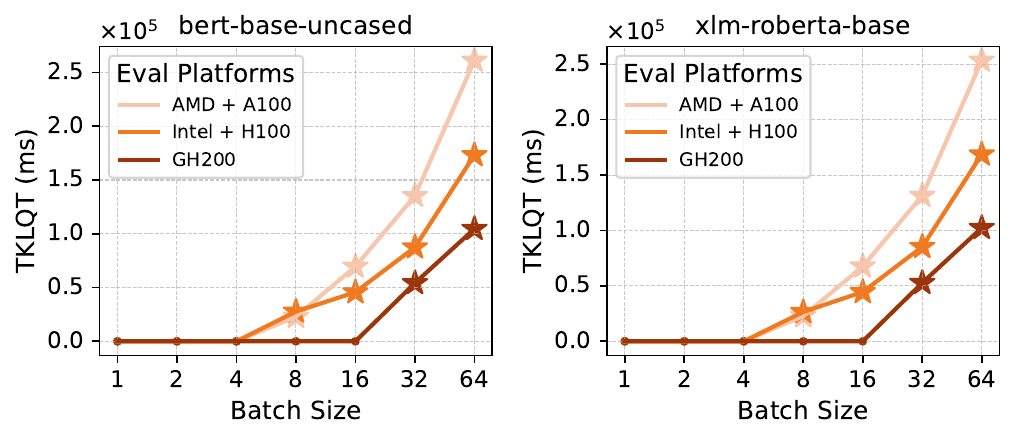}
    \caption{TKLQT values for encoder models Google-BERT and XLM-Roberta forward passes with increasing batch sizes on three evaluation platforms. Input sequence length is consistent at 512. The star marker indicates the approximate batch size where the workload transitions from CPU-bound (launch-dominated) to GPU-bound (queue-dominated).}
    \label{fig:encoder_classification}
\end{figure}

\subsection{Identifying Kernel Fusion Potential using Proximity Scores}
\label{sec:kernel_fusion}
Kernel fusion presents an optimization strategy to optimize inference latency by minimizing the kernel tax from launching PyTorch Aten operators from the CPU to the start of GPU kernel execution. Latency optimizations from kernel tax savings by kernel fusions can be relevant only to CPU-bound models, where the GPU is underutilized, and no kernel queuing takes place. We introduce a general kernel fusion method that recommends candidates based on kernel fusion chain length $L$. We use SKIP on two CPU-bound models, GPT2 and XLM-Roberta-Base, to identify prefill latency optimization opportunities, with Fig. \ref{fig:kernel_fusion_metrics} illustrating the distributions of unique kernel fusion candidates, the total instances of these candidates, and actual deterministic kernels candidates that can be fused during inference. While total instances of fusion candidates indicate the prevalence of fusion opportunities, the speedup is directly tied to the number of actual kernel chains fused, which are non-overlapping and with $PS=1$. Note that the speedups are idealized, with the theoretical maximum based solely on reducing the number of kernel launches, assuming constant launch overhead per kernel and no other performance impacts (positive or negative) from fusion. The reduction in kernel launches post-fusion is calculated as -
\begin{equation}
\label{eq:kfuse}
K_{fused} = K_{eager} - C_{fused} \times (L-1)
\end{equation}
where $K_{fused}$, $K_{eager}$ are the number of kernel launches post-kernel-fusion and during the eager mode, respectively, and $C_{fused}$ is the number of deterministic kernel fusion chains. Hence, the ideal speedup obtained through pure kernel launch savings is 
\begin{equation}
    Speedup = \frac{K_{eager}}{K_{fused}}
\end{equation}
with the speedup results reported in Fig. \ref{fig:fusion_speedup}. For both models in CPU-bound region, shorter chain lengths exhibit higher unique kernel fusion candidates (irrespective of PS) and total instances due to the greater likelihood of repeated kernel sequences. However, the speedup is modest (1.05x to 1.09x) since the chain length and the actual chains fused by the kernel with $PS=1$ are smaller. As $L$ increases, the number of unique candidates stabilizes while the total number of instances decreases. For longer chains (e.g., 64, 128, 256), the actual fusions are limited to a few non-overlapping chains. However, the idealized speedup becomes significant (up to 2.7x for GPT2 and 6.8x for XLM-Roberta-Base models) due to increased kernels fused and deployed on a single launch. Fusion effectiveness plateaus at chain lengths that exceed the eager mode kernel launches. 
We demonstrate scalable kernel fusion is a highly effective strategy for minimizing kernel launch overheads. 

\begin{figure}[t!]
    \centering
    \subfloat[Heatmap illustration of the distribution of unique kernel fusion chains detected for varying batch sizes for different chain lengths.]{\label{fig:encoder_runtime}
       	\includegraphics[width=\columnwidth]{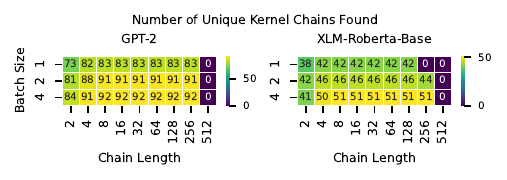}}\\ 
    \subfloat[Total instances of the chains detected for varying batch sizes for different chain lengths.]{\label{fig:encoder_gpu_idle_time}
       	\includegraphics[width=\columnwidth, height=3.4cm]{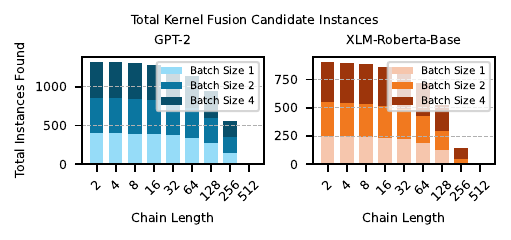}}\\
    \subfloat[Number of kernels fused with a proximity score of 1, for various chain lengths.]{\label{fig:encoder_cpu_idle_time}       	\includegraphics[width=\columnwidth]{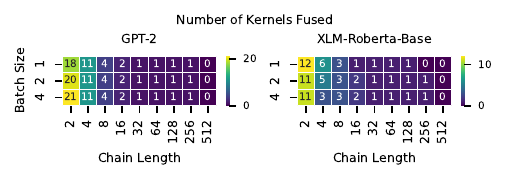}}\\
    \subfloat[Number of kernels launched during eager mode ($K_{eager}$) for inference prefill, for varying batch sizes. ]{\label{fig:encoder_cpu_idle_time}       	\includegraphics[width=\columnwidth, height=3 cm]{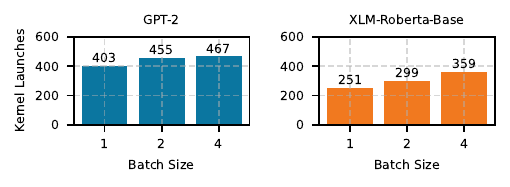}}
    \caption{Scalable kernel fusion recommendation metrics reported from SKIP during prefill inference on Intel+H100.
    }
    \label{fig:kernel_fusion_metrics}
\end{figure}
\begin{figure}[]
    \centering
    \includegraphics[width=0.5\textwidth, height=3.6cm]{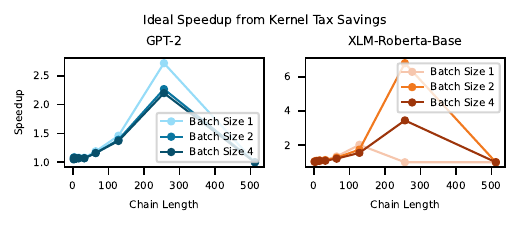}
    \caption{The potential ideal speedup realized purely from saving kernel launches by performing kernel fusions with varying chain lengths for inference prefill for the GPT2 and XLM-Roberta-Base models on the Intel+H100 platform. We assume (ideally) kernels are launched with the same launch latencies.}
    \label{fig:fusion_speedup}
\end{figure}

Note that the speedup calculated via Eq. \ref{eq:kfuse} and visualized for the proximity score-based fusion (blue bars in Fig. \ref{fig:speedup_comparison}) represents an idealized scenario. The orange bar represents torch.compile with reduce-overhead mode (utilizing only CUDA Graphs and not Triton backend compiler). For GPT-2 prefill and BS=1, the proximity score provides a best of 1.3x idealized speedup over the torch.compile for a chain length of 256 kernels. This calculation quantifies the theoretical maximum benefit achievable solely from minimizing the number of kernel launches and inherently assumes a constant launch overhead per kernel. This optimization strategy can yield significant benefits primarily in the CPU-bound region, where reducing kernel launch latency directly translates to lower overall inference latency and better overall PU utilization. In the GPU-bound region, launch overheads are often hidden or amortized by long-running kernels, making fusion for launch reduction less impactful.

\begin{figure}[]
    \centering
    \includegraphics[width=0.45\textwidth, height=3.2cm]{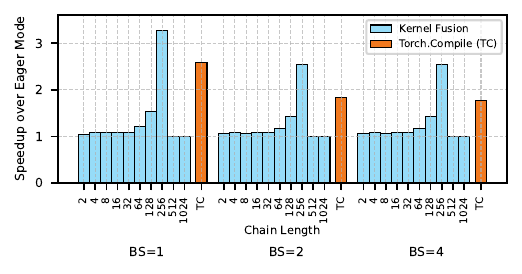}
    \caption{Relative to eager execution (no fusion), ideal kernel fusion speedups in blue bars 
    and torch.compile (TC) speedup in orange bar 
    for GPT-2 prefill on Intel+H100 are shown here.}
    \label{fig:speedup_comparison}
\end{figure}

\textbf{Key Takeaway: \textit{SKIP's kernel fusion recommendations can minimize kernel launches and improve inference performance, especially for CPU-bound models. Shorter chains yield modest speedups, while longer chains can maximize efficiency, especially for CC platforms like GH200.}}

\subsection{Platform Performance Analysis and Grace Case-Study}
\label{sec:grace}

We utilize SKIP to survey PU performance between LC systems (Intel+H100, AMD+A100) and the CC GH200, using encoder-only and decoder-only LLMs as our benchmark workloads for experimental evaluation. TTFT as inference latency is measured across varied batch sizes (BS) to assess performance dynamics, illustrated in Figs. \ref{fig:cutoff_encoder} and \ref{fig:cutoff_decoder}. Analyzing TTFT (Figs. \ref{fig:encoder_runtime} and  \ref{fig:decoder_runtime}), we observe clear performance crossover points (CPs). For encoder-only models, beyond the CP of BS=16, GH200 significantly reduces TTFT for large batch sizes (BS$>$16) compared to LC systems, providing a 1.6x/2.4x at BS=64 for Bert-Base-Uncased over Intel+H100/AMD+A100 systems. Similarly, CP is reached at BS=4 and BS=1 for decoder-only models of GP2 and Llama-3.2-1B, respectively, highlighting workload-specific sensitivities. For Llama-3.2-1B, speedup is 1.9x/2.7x at BS=16 over Intel+H100/AMD+A100. Batch sizes above the CP are the GPU-bound region, where the GH200 expectedly provides speedup for large-batched inference.

However, below the CP of BS$<$16 (CPU-bound region), the GH200 contrastingly consumes the highest latencies among all platforms. It exhibits 2.8x/1.9x and 1.7x/1.5x more latency for BS=1 and BS=8, respectively, for Bert-Base-Uncased over Intel+H100 and AMD+A100. The LC Intel+H100 consumes the least latency for small batch sizes. After the CP, the TTFT for the LC systems starts scaling linearly while the GH200 sustains near constant TTFT till BS=32 (4x more batch size compared to LC) before scaling linearly as its H100 GPU is utilized. Therefore, before the CP, the performance driver is CPU performance as (i) from TKLQT values in Fig. \ref{fig:encoder_classification}, GH200 before batch size 32 is CPU-bound and for LC systems, before batch size of 8, and (ii) H100 GPU is common for PCIe-connected H100 and GH200 systems. This indicates that CPU-bound LLM models (those with significant GPU idle time) do not see material benefits from the lower kernel launch latency of CC systems, and are primarily bound by the performance of the CPU. In this scenario, the Grace CPU comprising Arm Neoverse cores in the GH200 demonstrates a greater TTFT than the x86 processors (AMD EPYC 7313 connected to A100 and 2P Intel Xeon Platinum 8468V connected to H100), likely due to relatively lower CPU performance and/or less advanced software stack optimization.

Examining PU utilization (Figs. \ref{fig:encoder_gpu_idle_time}, \ref{fig:encoder_cpu_idle_time} for encoders; Figs. \ref{fig:decoder_gpu_idle_time}, \ref{fig:decoder_cpu_idle_time} for decoders) further establishes the reasoning above. The GH200 sustains CPU-bound operation longer, indicated by later inflection points (IPs) where the GPU saturates (GPU idle minimizes) and the CPU starts waiting (CPU idle increases). Consequently, GH200 achieves a balanced region of good CPU and GPU utilization at significantly higher batch sizes (encoders: LC BS=4-8, CC BS=16-32; decoders: LC BS=2-4, CC BS=4-8). This demonstrates the potential of CC architectures to maintain system balance on larger scales, enabled by efficient memory subsystems. Conversely, the GH200's higher low-batch latency underscores the importance of CPU performance in the launch-dominated region relative to the x86 CPUs in the LC systems. This bottleneck highlights the value of optimization strategies like kernel fusion (discussed in Sec. \ref{sec:kernel_fusion}), which can mitigate the launch tax and improve GH200's performance in latency-critical, low-batch scenarios, and move towards balanced PU-resource utilization. Hence, the potential for scalable kernel fusion optimization is more for CC/TC architectures. In the case of higher model complexity, like  Llama-3.2-1B model (Fig. \ref{fig:cutoff_decoder}), while there is no CP (latency is similar at the batch size of 1), there is significant GPU idle time to lower inference latency through kernel fusion.

\begin{figure}[]
    \centering
    \subfloat[Inference time for the two encoder models with varying batch sizes.]{\label{fig:encoder_runtime}
       	\includegraphics[width=\columnwidth]{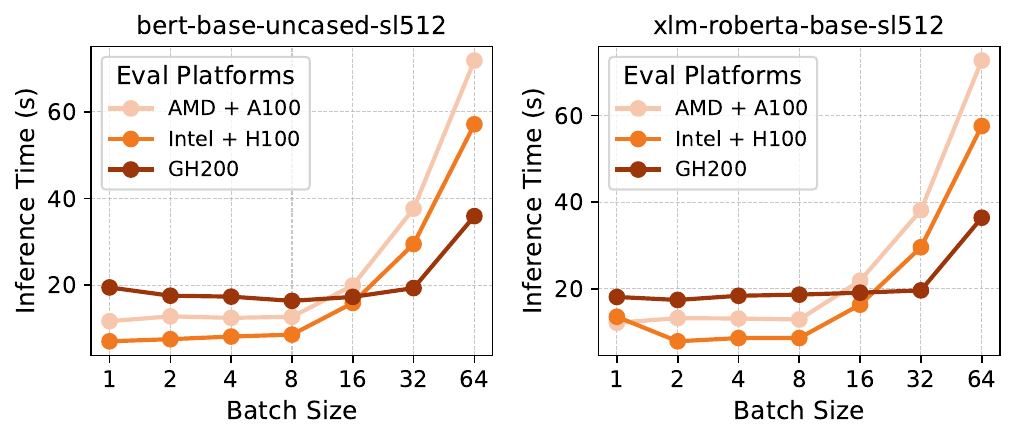}}\\
    \subfloat[GPU idle time for the two encoder models with varying batch sizes.]{\label{fig:encoder_gpu_idle_time}
       	\includegraphics[width=\columnwidth]{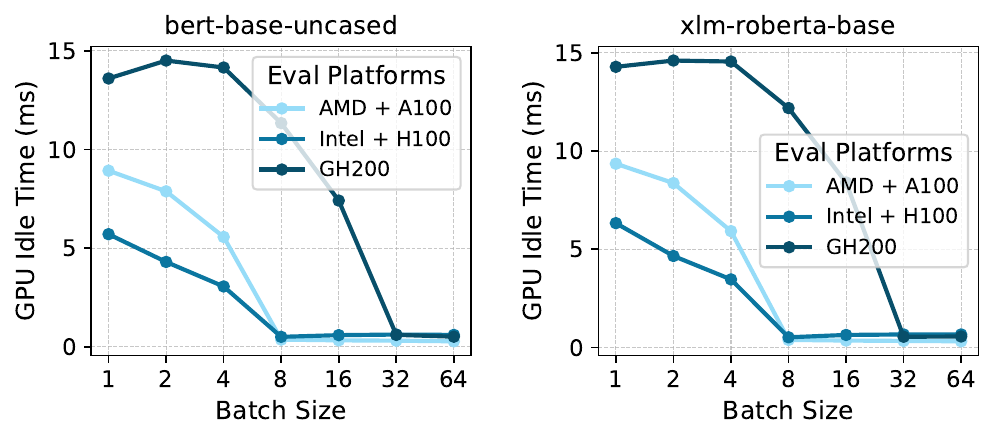}}\\
    \subfloat[CPU idle time for the two encoder models with varying batch sizes.]{\label{fig:encoder_cpu_idle_time}       	\includegraphics[width=\columnwidth]{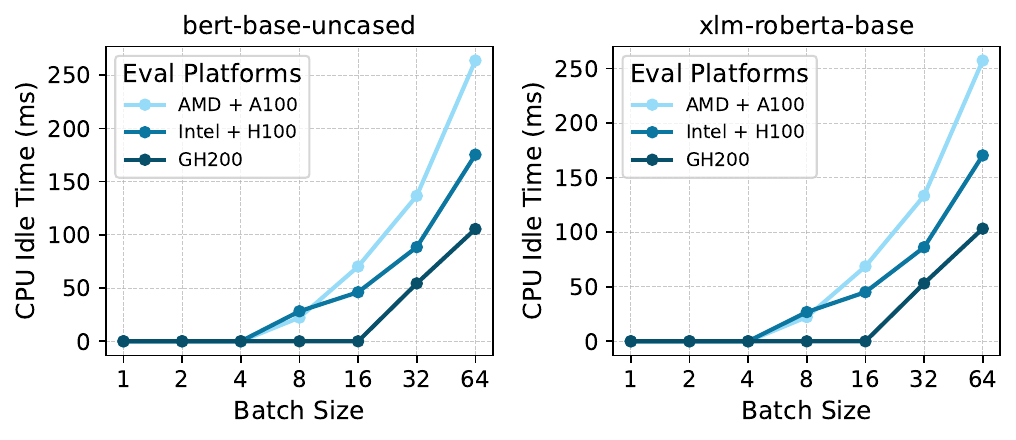}}\\
    \caption{Prefill inference latency, GPU, and CPU idle times for encoder models Bert-Base-Uncased and XLM-Roberta-Base on the three evaluation platforms.
    }
    \label{fig:cutoff_encoder}
\end{figure}

\begin{figure}[ht!]
    \centering
    \subfloat[Inference time for the two decoder models with varying batch sizes.]{\label{fig:decoder_runtime}
       	\includegraphics[width=\columnwidth]{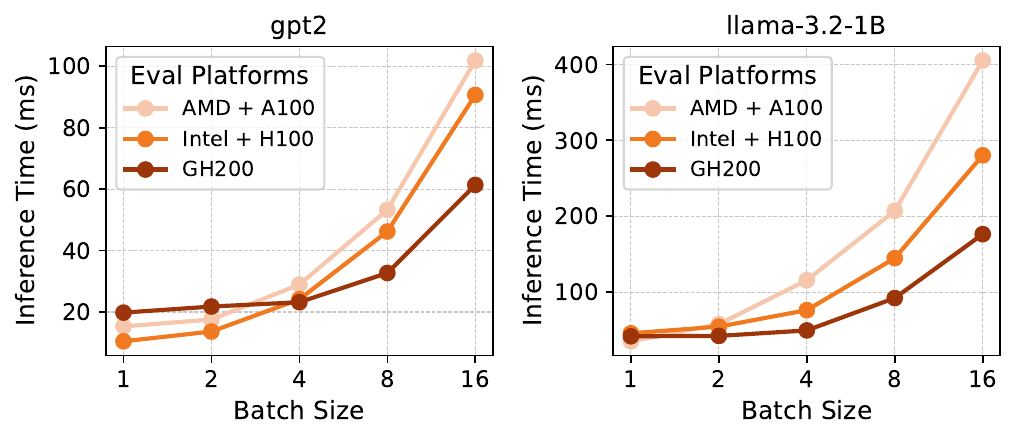}}\\
    \subfloat[GPU idle time for the two decoder models with varying batch sizes.]{\label{fig:decoder_gpu_idle_time}
       	\includegraphics[width=\columnwidth]{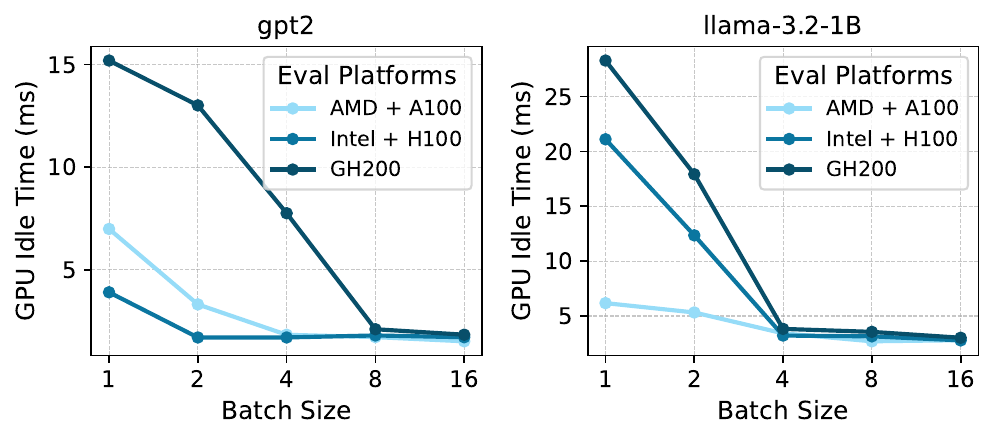}}\\
    \subfloat[CPU idle time for the two decoder models with varying batch sizes.]{\label{fig:decoder_cpu_idle_time}       	\includegraphics[width=\columnwidth]{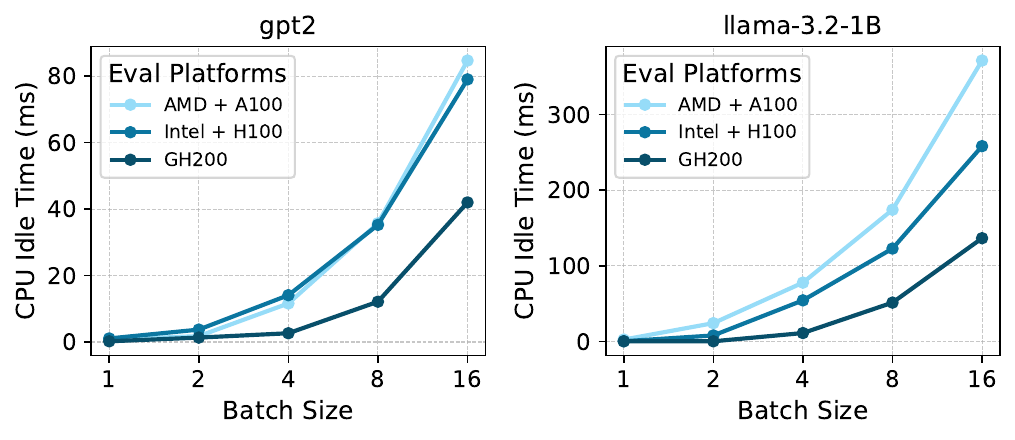}}\\
    \caption{
    Prefill inference latency, GPU, and CPU idle times for decoder models GPT2 and Llama-3.2-1B on the three evaluation platforms.
    }
    \label{fig:cutoff_decoder}
\end{figure}

\textbf{Key Takeaway: \textit{GH200's high bandwidth memory and interconnect make the GPU well utilized and push the bottleneck to the CPU across a wider range of batch sizes. However, Grace CPU's single-thread performance constrains GH200 in CPU-bound workloads for small batch sizes and latency-sensitive applications. Addressing these bottlenecks requires enhancing CPU performance or employing intelligent scheduling in CC/TC designs.}}

\section{Conclusions \& Future Works}

In this work, we demonstrate that CC architectures like GH200 offer superior inference at larger batch sizes due to high-bandwidth memory and unified virtual memory. From our observations, there are three main contributors to inference latency optimization - GPU performance, CPU performance, and type of coupling paradigm. We identify through benchmarking that compared to LC architectures, the CC systems remain CPU-bound at larger batch sizes, where CPU performance is more critical. Due to the single-threaded performance of the CPU relative to the CPUs on LC systems, inference time is longer in GH200 for smaller batch sizes. 

We demonstrate one way to mitigate the CPU limitation of CC architecture by proposing a novel kernel fusion method based on proximity scores to find deterministic chains to get speedup by minimizing kernel launches. Idealized potential prefill inference speedups of up to 2.7x for GPT2 and up to 6.8x for XLM-Roberta-Base models can be achieved through general kernel fusions on Intel+H100  (Fig. \ref{fig:fusion_speedup}) by minimizing kernel launches. We can extend kernel fusion optimization to higher batch sizes for GH200, compensating for the lower performance of the Grace CPU. Our novel profiling tool, SKIP, and the proposed kernel-level performance metrics enabled these insights, allowing for fine-grained analysis of operator-kernel dynamics and accurate characterization of CPU vs. GPU boundedness across platforms.

This work is an initial effort at experimental evaluation and characterization of LLM workloads on physical systems with varying degrees of CPU-GPU coupling.
We plan to extend our experimental analysis to include systems like NVIDIA GB200 and AMD MI300A for future work. These systems were not available in time for inclusion in this work. 
We also plan to broaden our workload scope to include recommendation models (RMs) and graph neural networks (GNNs). We also plan to implement a more comprehensive kernel fusion prototype to validate the predicted performance gains across different platforms and models. More experimentation is needed to fully understand the interaction between modern AI workloads and evolving heterogeneous computing systems.

\bibliographystyle{plain}
\bibliography{references}

\end{document}